\documentclass[12pt,english]{article}
\usepackage[T1]{fontenc}
\usepackage[latin9]{inputenc}
\usepackage{textcomp}
\usepackage{amsmath}
\usepackage{graphicx}
\usepackage{amssymb}
\usepackage{esint}

\makeatletter

\DeclareRobustCommand{\greektext}{%
  \fontencoding{LGR}\selectfont\def\encodingdefault{LGR}}
\DeclareRobustCommand{\textgreek}[1]{\leavevmode{\greektext #1}}
\DeclareFontEncoding{LGR}{}{}
\DeclareTextSymbol{\~}{LGR}{126}
\newcommand{\lyxmathsym}[1]{\ifmmode\begingroup\def\b@ld{bold}
  \text{\ifx\math@version\b@ld\bfseries\fi#1}\endgroup\else#1\fi}

\providecommand{\tabularnewline}{\\}

\newcommand{\lyxaddress}[1]{
\par {\raggedright #1
\vspace{1.4em}
\noindent\par}
}

\makeatother

\usepackage{babel}

\begin{document}

\title{Tentative Structural Features of a Gapped RVB State in the Anisotropic
Triangular Lattice.}

\author{A.L. Tchougréeff$^{a,b,c}$ and R. Dronskowski$^{a}$}

\maketitle

\lyxaddress{$^{a}$Institut für anorganische Chemie, RWTH - Aachen University,
Landoltweg 1, D-52056 Aachen, Germany;\\
$^{b}$Poncelet Lab., Independent University of Moscow, Moscow
Center for Continuous Mathematical Education, Moscow, Russia;\\
$^{c}$Division of Electrochemistry, Department of Chemistry, Moscow
State University, Moscow, Russia. }
\begin{abstract}
The self-consistency equations for the independent order parameters
as well as the free energy expression for the mean-field RVB model
of the spin-1/2 Heisenberg Hamiltonian on the anisotropic triangular
lattice is considered in the quasi-one-dimensional approximation.
The solutions of the self-consistency equations in the zero-temperature
limit are in fair agreement with the previous numerical analysis of
the same model by other authors. In particular, the transition from
the ungapped 1D-RVB state to the gapped 2D-RVB state occurs at an
arbitrarily weak transversal exchange ($J_{2}\rightarrow0)$ although
the amount of the gap is exponentially small: $\frac{12J_{1}}{\pi}\exp\left(-\frac{2J_{1}}{J_{2}}\right)$,
where $J_{1}$ is the longitudinal exchange parameter. The structural
consequences of the formation of the 2D-RVB state are formulated by
extending the famous bond order \emph{vs}. bond length relation known
for polyenes (one-dimensional Hubbard chains). Analytical estimates
of this effect are given. 
\end{abstract}

\section{Introduction}

\begin{figure}
\center{\includegraphics[scale=0.6]{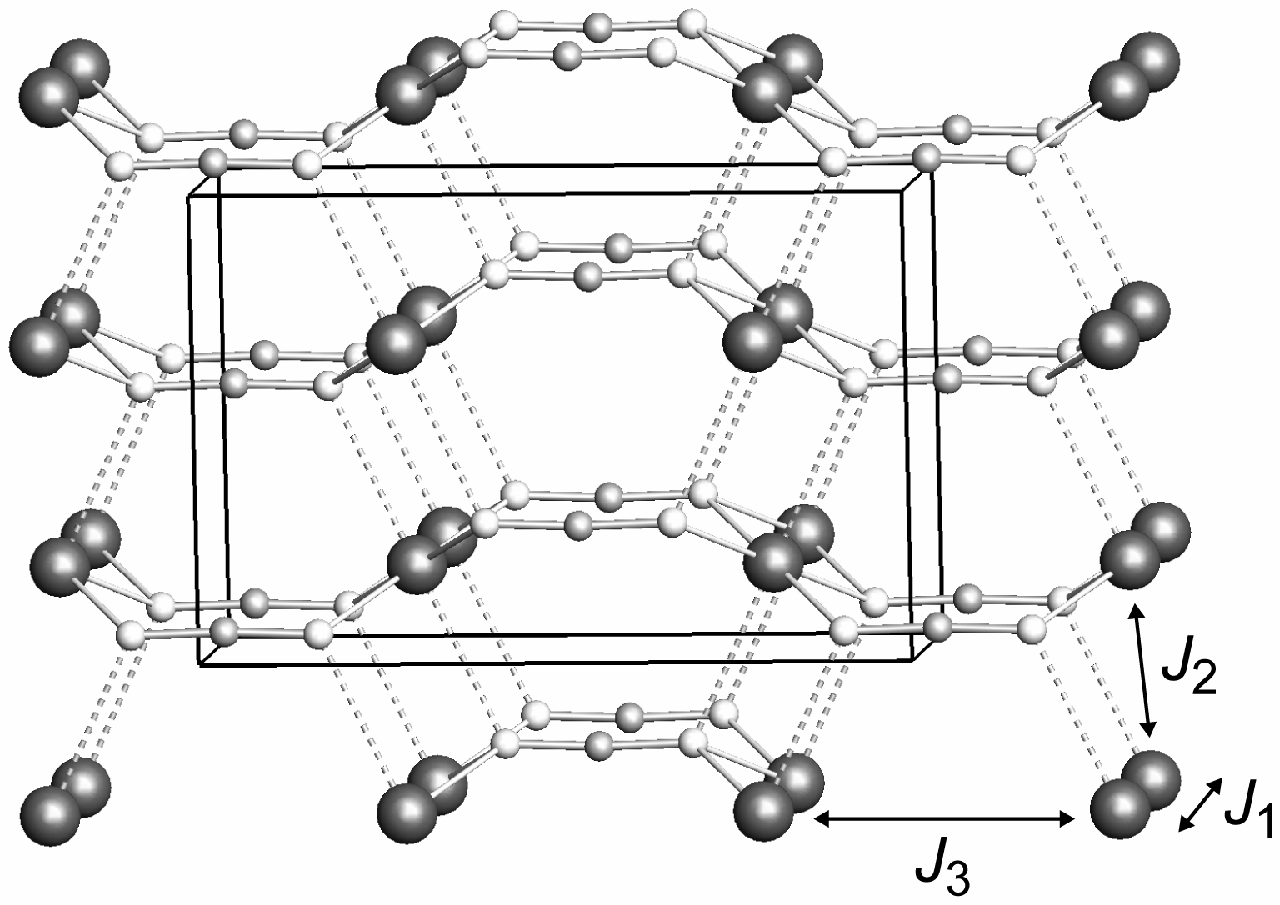}\\
 a\\
 \includegraphics[scale=0.5]{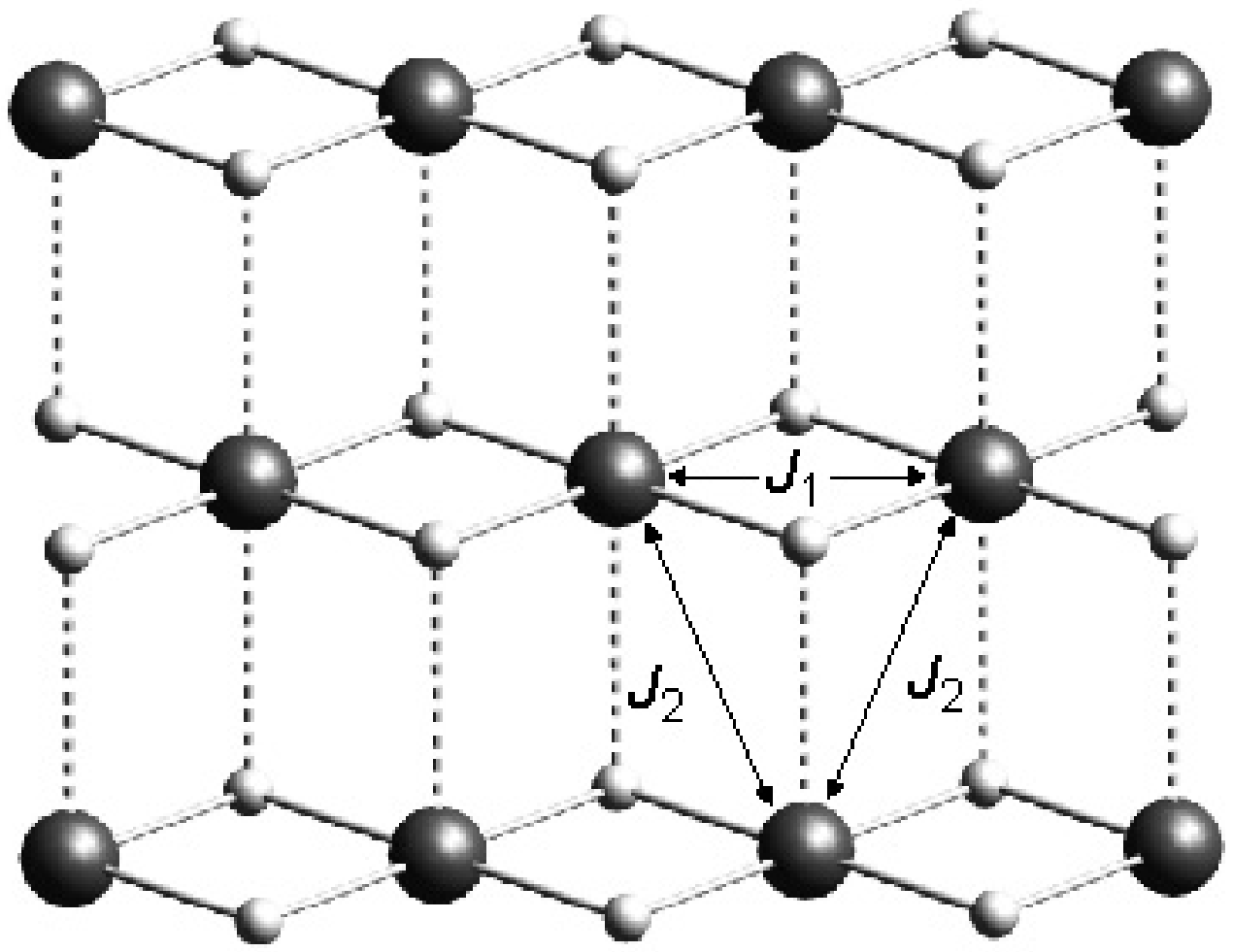}\\
 b} \caption{A look into the CuNCN crystal structure. A stronger $J_{1}$ extends
in the $a$ direction; a somewhat weaker $J_{2}$ extends along the
$b\pm a$ directions. The weakest $J_{3}$ extends in the $c$ direction
and is not considered in the present paper.}

\label{fig:CuNCN-ab-layer} %
\end{figure}

The RVB state originally proposed by Pauling \cite{Pauling} for describing
the structure of the benzene molecule is being sought in many materials
after Anderson's \cite{Anderson} conjecture that it represents the
ground state of cuprate-based high-temperature superconductors. The
recently obtained \cite{Dronskowski} CuNCN phase whose structure
is represented in Fig. \ref{fig:CuNCN-ab-layer} had been proposed
as a candidate for an RVB ground state spin liquid \cite{Tchougreeff-Dronskowski-Arxiv }
due to frustration of the effective exchange in the \emph{ab}-plane
where the 1/2 Cu$^{2+}$ local spins form an anisotropic triangular
lattice. The material had been subject of a series of measurements
of its magnetic susceptibility, electric resistivity, heat capacity,
also elastic neutron scattering as well as of ESR, NMR relaxation,
and muon spin resonance (all \emph{vs.} $T$) \cite{Dronskowski,NoMagneticScattering,10SmallNegroes}.
Although the neutron scattering (complete absence of the magnetic
signal) as well as the susceptibility measurements (temperature independent
paramagnetism above 80 K changed to approximately activation decay
of the susceptibility below this temperature) strongly indicate the
transition between the 1D-RVB regime at higher temperatures to the
gapped 2D-RVB regime below 80 K the issue remains controversial \cite{10SmallNegroes}
since the NMR and the $\mu$SR data so far better fit into a model
of inhomogeneous spin-glass-like ground state. %
In order to solve this dilemma we try to find out in the present paper
what might be the structural concequences of the formation of two
spin-liquid phases which is going to help in obtaining direct experimental
evidence of this picture.

\section{RVB mean-field analysis of an anisotropic triangular lattice system}

\subsection{Hamiltonian}

A close inspection of the structure (Fig. \ref{fig:CuNCN-ab-layer})
reveals that each Cu$^{2+}$ ion can be effectively antiferromagnetically
coupled to two of its neighbors forming a chain while somewhat weaker
antiferromagnetic coupling with four more neighbors from two adjacent
parallel chains results in a Heisenberg model on an anisotropic triangular
lattice with the Hamiltonian: \begin{equation}
\sum_{\mathbf{r}}\sum_{\mathbf{\tau}}J_{\mathbf{\boldsymbol{\tau}}}\mathbf{S}_{\mathbf{r}}\mathbf{S}_{\mathbf{r}+\tau}\label{eq:Hamiltonian}\end{equation}
where the coupling vectors $\mathbf{\tau}$ take three values $\tau_{i};\, i=1\div3;\,\tau_{1}=(1,0);\,\tau_{2}=(\frac{1}{2},\frac{\sqrt{3}}{2});\,\tau_{3}=(\frac{1}{2},-\frac{\sqrt{3}}{2})$
with the interaction of the strength $J_{1}$ along the lattice vector
$\mathbf{\mathbf{\tau}_{\mathrm{1}}}$ (two neighbors) and with a
somewhat smaller strength $J_{2}$ along the lattice vectors $\mathbf{\mathbf{\tau}_{\mathrm{2}}}$
and $\mathbf{\mathbf{\tau}_{\mathrm{3}}}$ (two neighbors along each).
This is precisely the setting for which Yunoki and Sorella \cite{YunokiSorella}
proposed that two (different) spin-singlet RVB (s-RVB) states are
formed at different temperatures depending on the amount of anisotropy
$\frac{J_{2}}{J_{1}}$ on the basis of their VMC calculations. Later
Hayashi and Ogata \cite{Hayashi and Ogata } reproduced this result
within a mean-field treatment which we basically follow here.

\subsection{Equations of motion and self consistency equations}

Hayashi and Ogata \cite{Hayashi and Ogata } base their analysis of
the Hamiltonian eq. (\ref{eq:Hamiltonian}) on returning to the electron
representation from the spin representation by the standard formulae:\begin{equation}
\mathbf{S}_{i}=\frac{1}{2}c_{i\alpha}^{+}\mathbf{\boldsymbol{\sigma}}_{\alpha\beta}c_{i\beta},\label{eq:SpinThroughFermi}\end{equation}
 where $c_{i\sigma}^{+}(c_{i\sigma})$ are the electron creation (annihilation)
operators subject to the Fermi anticommutation relations; $\mathbf{\boldsymbol{\sigma}}_{\alpha\beta}$
are the elements of the Pauli matrices and the summation over repeating
indices is assumed. For the latter one can derive equations of motion
based on the Heisenberg representation in which each operator obeys
the following equation of motion:\begin{equation}
i\hbar\dot{A}=\left[A,H\right]\label{eq:HeisenbergEOM}\end{equation}
 where $\left[,\right]$ stands for the commutator of the operators
and the dot-on-top symbol for the time derivative. Applying this to
the creation and annihilation operators $c_{\mathbf{r}\sigma}^{+}(c_{\mathbf{r}\sigma})$
and performing commutation, mean-field decoupling and Fourier transformation
as done in \cite{Tchougreeff-Dronskowski-Arxiv ,Hayashi and Ogata }
results in mean-field equations of motion for these operators : \begin{eqnarray}
i\hbar\dot{c}_{\mathbf{k}\sigma}=-\frac{3}{2}\sum_{\mathbf{\tau}}J_{\mathbf{\boldsymbol{\tau}}}\xi_{\mathbf{\boldsymbol{\tau}}}\cos(\mathbf{k\tau})c_{\mathbf{k}\sigma} & -\frac{3}{2}\sum_{\mathbf{\tau}}J_{\mathbf{\boldsymbol{\tau}}}\Delta_{\boldsymbol{\tau}} & \cos(\mathbf{k\tau})c_{-\mathbf{k}-\sigma}^{+}\nonumber \\
i\hbar\dot{c}_{\mathbf{k}\sigma}^{+}=\frac{3}{2}\sum_{\mathbf{\tau}}J_{\mathbf{\boldsymbol{\tau}}}\xi_{\mathbf{\boldsymbol{\tau}}}\cos(\mathbf{k\tau})c_{\mathbf{k}\beta}^{+} & +\frac{3}{2}\sum_{\mathbf{\tau}}J_{\mathbf{\boldsymbol{\tau}}}\Delta_{\boldsymbol{\tau}}^{*} & \cos(\mathbf{k\tau})c_{-\mathbf{k}-\sigma}\label{eq:MeanFieldEOM}\end{eqnarray}
These reduce to the set of $2\times2$ eigenvalue problems for each
wave vector $\boldsymbol{\mathbf{k}}$:

\begin{equation}
\left(\begin{array}{cc}
\xi_{\mathbf{k}} & \Delta_{\mathbf{k}}\\
\Delta_{\mathbf{k}}^{*} & -\xi_{\mathbf{k}}\end{array}\right)\left(\begin{array}{c}
u_{\mathbf{k}}\\
v_{\mathbf{k}}\end{array}\right)=E_{\mathbf{k}}\left(\begin{array}{c}
u_{\mathbf{k}}\\
v_{\mathbf{k}}\end{array}\right)\label{eq:BogolyubovEquation}\end{equation}
with\begin{eqnarray}
\xi_{\mathbf{k}} & = & -3\sum_{\mathbf{\tau}}J_{\mathbf{\boldsymbol{\tau}}}\xi_{\mathbf{\boldsymbol{\tau}}}\cos(\mathbf{k\tau})\nonumber \\
\Delta_{\mathbf{k}} & = & 3\sum_{\mathbf{\tau}}J_{\mathbf{\boldsymbol{\tau}}}\Delta_{\mathbf{\boldsymbol{\tau}}}\cos(\mathbf{k\tau})\label{eq:DispersionFunctions}\end{eqnarray}
(summation over $\mathbf{\boldsymbol{\tau}}$ extends to $\pm\tau_{i};i=1\div3$)
which results in the eigenvalues (quasiparticle spectrum) of the form:

\begin{eqnarray}
E_{\mathbf{k}} & = & \sqrt{\xi_{\mathbf{k}}^{2}+\left|\Delta_{\mathbf{k}}\right|^{2}}\label{eq:eigenvalues}\end{eqnarray}
whose eigenvectors are combinations of the destruction and creation
operators with the above Bogoliubov transformation coefficients $u_{\mathbf{k}},v_{\mathbf{k}}$.
These equations result in the self-consistency equations of the form:\begin{eqnarray}
\xi_{\mathbf{\boldsymbol{\tau}}} & = & -\frac{1}{2N}\sum_{\mathbf{k}}\exp(i\mathbf{k\tau})\frac{\xi_{\mathbf{k}}}{E_{\mathbf{k}}}\tanh\left(\frac{E_{\mathbf{k}}}{2\theta}\right)\nonumber \\
\Delta_{\mathbf{\boldsymbol{\tau}}} & = & \frac{1}{2N}\sum_{\mathbf{k}}\exp(-i\mathbf{k\tau})\frac{\Delta_{\mathbf{k}}}{E_{\mathbf{k}}}\tanh\left(\frac{E_{\mathbf{k}}}{2\theta}\right)\label{eq:SelfConsistencyEquations}\end{eqnarray}
for six order parameters $\xi_{\mathbf{\boldsymbol{\tau}}},\Delta_{\mathbf{\boldsymbol{\tau}}}$
defined as 

\begin{eqnarray}
\xi_{\mathbf{\boldsymbol{\tau}}} & = & \left\langle c_{\mathbf{r}+\boldsymbol{\tau}\sigma}^{+}c_{\mathbf{r}\sigma}\right\rangle \nonumber \\
\Delta_{\boldsymbol{\tau}} & = & \left\langle c_{\mathbf{r}\alpha}c_{\mathbf{r}+\boldsymbol{\tau}\beta}\right\rangle .\label{eq:OrderParameters}\end{eqnarray}
It is remarkable and important for the subsequent treatment that the
order parameters $\xi_{\mathbf{\boldsymbol{\tau}}}$ are in fact \emph{bond
orders} for the corresponding pairs of atoms.

\subsection{Free energy}

Following Ref. \cite{Ogata&Fukuyama} one can write immediately the
free energy in terms of the above order parameters: \begin{equation}
F(\theta)=-\frac{\theta}{2N}\sum_{\mathbf{k}}\ln\left(2\cosh\left(\frac{E_{\mathbf{k}}}{2\theta}\right)\right)+\frac{3}{2}\sum_{\mathbf{\tau}}J_{\mathbf{\boldsymbol{\tau}}}\xi_{\mathbf{\boldsymbol{\tau}}}^{2}+\frac{3}{2}\sum_{\mathbf{\tau}}J_{\mathbf{\boldsymbol{\tau}}}\left|\Delta_{\mathbf{\mathbf{\boldsymbol{\tau}}}}\right|^{2}\label{eq:FreeEnergy}\end{equation}
where $\theta=k_{B}T$ and summation over $\mathbf{\boldsymbol{\tau}}$
extends to $\pm\tau_{i};i=1\div3$. Minima of this expression with
respect to $\xi_{\mathbf{\boldsymbol{\tau}}}$ and $\Delta_{\mathbf{\boldsymbol{\tau}}}$
correspond to various possible states of the system.

\section{Simplified RVB on the anisotropic triangular lattice}

\noindent The numerical analysis \cite{Hayashi and Ogata } shows
that in agreement with general theorems \cite{SU2-symmetry} the order
parameters satisfy additional phase relations 

\begin{equation}
\arg\Delta_{\boldsymbol{\tau}_{2}}-\arg\Delta_{\boldsymbol{\tau}_{3}}=\pm\frac{\pi}{2}\label{eq:PhaseRelations}\end{equation}

\noindent which allows to reduce the number of order parameters to
only two \cite{Hayashi&Ogata-arxiv,Tchougreeff-Dronskowski-Arxiv }:\begin{equation}
\sqrt{2}\xi=\xi_{\boldsymbol{\tau}_{1}};\sqrt{2}\eta=\left|\Delta_{\boldsymbol{\tau}_{2}}\right|=\left|\Delta_{\boldsymbol{\tau}_{3}}\right|\label{eq:SymmetricOrderParameters}\end{equation}
the first responsible for establishing the gapless 1D-RVB state within
the chains and the second one for opening a gap and establishing the
2D-RVB in the transversal direction and other three order parameters
set to be zero. It can be shown that this particular choice of the
phases leads to the following form of the fermion quasiparticle spectrum: 

\begin{eqnarray}
E_{\mathbf{k}}^{2} & = & 18\left(J_{1}^{2}\xi^{2}\cos^{2}(\mathbf{k}_{x})+J_{2}^{2}\eta^{2}+J_{2}^{2}\eta^{2}\cos(\mathbf{k}_{x})\cos(\mathbf{k}_{y}\sqrt{3})\right).\label{eq:Spectrum}\end{eqnarray}
This form of the quasiparticle spectrum allows for a simple analysis.
Obviously in the 1D-RVB state ($\eta=0$) the spectrum has a vanishing
dispersion in the $y$-direction:

\begin{eqnarray}
E_{\mathbf{k}} & =3\sqrt{2} & \left|J\xi\cos(\mathbf{k}_{x}).\right|\label{eq:LongitudinalSpectrum}\end{eqnarray}
It is gapless along the lines $\mathbf{k}_{x}=\pm\frac{\pi}{2}$ and
contains ridges at $\mathbf{k}_{x}=0,\pm\pi$ extended in the $y$-direction.
The corresponding density of the quasiparticle states (qDOS) is depicted
in Fig. 2 (red). The divergence on the qDOS:

\[
g(\varepsilon)=\frac{2}{\pi\sqrt{18J^{2}\xi^{2}-\varepsilon^{2}}}\]
on the upper border of the spectrum is obviously due to the abovementioned
ridges in the spectrum of the quasiparticles. On the other hand the
low-energy behavior of the system in the 1D-RVB state is controlled
by a constant qDOS at the zero energy, which is perfectly reflected
in the temperature independent paramagnetic susceptibility in the
respective temperature region. 

\begin{figure}
\center{ \includegraphics{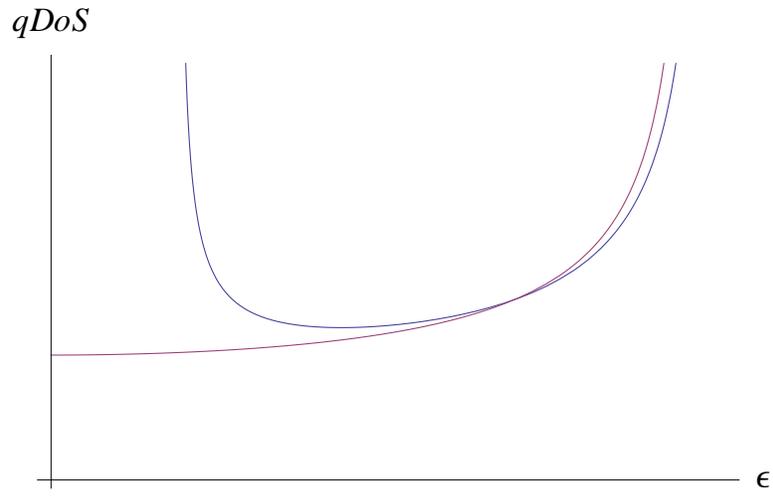}} \label{fig:qDoS}\caption{The schematic densities of the fermion quasiparticle states in the
1D-RVB (red) and 2D-RVB states (blue) of a system of 1/2 Heisenberg
spins on the anisotropic triangular lattice. The 2D-RVB qDOS is built
for the effective anisotropy parameter $a=\frac{1}{5}$ (see the text
for details). In addition to the gap opening one may observe a slight
shift of the energy at which the upper singularity occurs. }
\end{figure}

The transition to the 2D-RVB state is accompanied by a significant
reorganization of the quasiparticle spectrum. Both the ridges at $\mathbf{k}_{x}=0,\pm\pi$
and the degeneration lines at $\mathbf{k}_{x}=\pm\frac{\pi}{2}$ disappear
and are replaced by the critical points whose characteristics are
given in Table \ref{Tab:CriticalPoints}. One can realize that in
the 2D case the qDOS is mainly contributed by the logarithmic van
Hove singularities due to the saddle points of the spectrum on top
of the constant contributions coming from the minima and maxima of
the quasiparticle spectrum. Physically it must be important that the
values of the energy corresponding to the minima and maxima differ
from the saddle point energies only in the second order in a small
parameter of efficient anisotropy: $a=\frac{J_{2}\eta}{J_{1}\xi}$
so that one ultimately cannot expect anything, but some widening of
the logarithmic peak of the qDOS on the upper bound of the spectrum.
The most important changes occur at the lower bound of the spectrum,
where a gap opens. As one can see from Table \ref{Tab:CriticalPoints}
the singular spectral weight must concentrate at the energy of the
lower saddle points: $3\sqrt{2}J_{2}\eta$, although the lower boundary
of the spectrum is smaller than this value in the fourth order with
respect to efficient anisotropy. 

\begin{table}
\caption{Critical points of the quasiparticle spectrum in the gapped spin liquid
2D-RVB state of a system of 1/2 Heisenberg spins on the anisotropic
triangular lattice. $(\mathbf{k}_{x},\mathbf{k}_{y})$ stands for
the coordinates of the critical point in the Brillouin zone, \emph{n
}is the degeneracy - total number of points of the given type; other
entries are self-explanatory. }
\label{Tab:CriticalPoints}\center{\begin{tabular}{|c|c|c|c|}
\hline 
$(\mathbf{k}_{x},\mathbf{k}_{y})$ & \emph{n} & $E^{2}$  & point type\tabularnewline
\hline
\hline 
$(0,\frac{\pi l}{\sqrt{3}});l=-2,0,2$  & 3  & $18\left(J_{1}^{2}\xi^{2}+2J_{2}^{2}\eta^{2}\right)$  & maximum\tabularnewline
\hline 
$(\pm\pi,\pm\frac{\pi}{\sqrt{3}})$  & 4  & $18\left(J_{1}^{2}\xi^{2}+2J_{2}^{2}\eta^{2}\right)$  & maximum\tabularnewline
\hline 
$(0,\pm\frac{\pi}{\sqrt{3}})$  & 2  & $18J_{1}^{2}\xi^{2}$  & saddle\tabularnewline
\hline 
$(\pm\pi,0)$  & 2  & $18J_{1}^{2}\xi^{2}$  & saddle\tabularnewline
\hline 
$(\pm\frac{\pi}{2},\pm\frac{\pi}{2\sqrt{3}})$  & 4  & $18J_{2}^{2}\eta^{2}$  & saddle\tabularnewline
\hline 
$(\pm\frac{\pi}{2},\pm\frac{\pi\sqrt{3}}{2})$  & 4  & $18J_{2}^{2}\eta^{2}$  & saddle\tabularnewline
\hline 
$(\pm\arccos\left(-\frac{J^{\prime2}\eta^{2}}{2J^{2}\xi^{2}}\right),\pm\frac{2\pi}{\sqrt{3}})$  & 4  & $18J_{2}^{2}\eta^{2}\left(1+\frac{3}{4}\frac{J_{2}^{2}\eta^{2}}{J_{1}^{2}\xi^{2}}\right)$  & saddle\tabularnewline
\hline 
$(\pm\arccos\left(\frac{J^{\prime2}\eta^{2}}{2J^{2}\xi^{2}}\right),\pm\frac{\pi}{\sqrt{3}})$  & 4  & $18J_{2}^{2}\eta^{2}\left(1-\frac{1}{4}\frac{J_{2}^{2}\eta^{2}}{J_{1}^{2}\xi^{2}}\right)$  & minimum\tabularnewline
\hline 
$(\pm\arccos\left(-\frac{J^{\prime2}\eta^{2}}{2J^{2}\xi^{2}}\right),0)$  & 2  & $18J_{2}^{2}\eta^{2}\left(1-\frac{1}{4}\frac{J_{2}^{2}\eta^{2}}{J_{1}^{2}\xi^{2}}\right)$  & minimum\tabularnewline
\hline
\end{tabular}}%
\end{table}
On the other hand we notice that the wave vectors of the critical
points of the quasiparticle spectrum in the vicinity of its lower
boundary either have the $\mathbf{k}_{x}$ component equal to $\pm\frac{\pi}{2}$
or differ from these values in a higher order of efficient anisotropy.
Following \cite{Hayashi&Ogata-arxiv} we notice that in this range
of wave vectors the dispersion of quasiparticles in the $y$-direction
is negligeably small. This brings us to the idea that one can hope
that neglecting the $y$-dispersion when calculating the integral
characteristics of the system does not affect the precision catastrophically.
This may be considered as a quasi-one-dimensional approximation for
the spectrum, which then takes the form:\begin{eqnarray}
E_{\mathbf{k}} & = & \sqrt{18\left(J_{1}^{2}\xi^{2}\cos^{2}(\mathbf{k}_{x})+J_{2}^{2}\eta^{2}\right)}.\label{eq:Q1DSpectrum}\end{eqnarray}
Inserting this in the standard definition of the density of states
we obtain as expected: \begin{equation}
g(\varepsilon)=\frac{2\varepsilon}{\pi\sqrt{\left(18J_{1}^{2}\xi^{2}+18J_{2}^{2}\eta^{2}-\varepsilon^{2}\right)\left(\varepsilon^{2}-18J_{2}^{2}\eta^{2}\right)}}\label{eq:Q1DqDOS}\end{equation}
which corresponds to the quasiparticle band ranging on the energy
scale from $3\sqrt{2}J_{2}\eta$ to $3\sqrt{2}\sqrt{J_{1}^{2}\xi^{2}+J_{2}^{2}\eta^{2}}$
with its lower boundary being as explained above somewhat higher than
the lower boundary of the exact spectrum and with the upper boundary
being located between the upper boundary and logarithmic peak of the
exact spectrum. 

With use of the qDOS eq. (\ref{eq:Q1DqDOS}) one can easily write
the explicit expression for the free energy as relying on the general
expression eq. (\ref{eq:FreeEnergy}). It reads as follows:

\begin{equation}
F=6J_{1}\xi^{2}+12J_{2}\eta^{2}-2\theta\intop g(\varepsilon)\ln\left(2\cosh\left(\frac{\varepsilon}{2\theta}\right)\right)d\varepsilon.\label{eq:Q1DFreeEnergy}\end{equation}
We do not expect that our results obtained in \cite{Tchougreeff-Dronskowski-Arxiv }
with use of the high-temperature expansion: \[
\ln\left(2\cosh\left(\frac{E_{\mathbf{k}}}{2\theta}\right)\right)\approx\ln2+\frac{1}{2}\left(\frac{E_{\mathbf{k}}}{2\theta}\right)^{2}-\frac{1}{12}\left(\frac{E_{\mathbf{k}}}{2\theta}\right)^{4}\]
which allowed to perform integration over the entire BZ without neglecting
the $y$-dispersion are going to change. By contrast, in the low temperature
regime we first rewrite 

\[
\theta\ln\left(2\cosh\left(\frac{\varepsilon}{2\theta}\right)\right)=\frac{\varepsilon}{2}+\theta\ln\left(1+\exp\left(-\frac{\varepsilon}{\theta}\right)\right)\]
and immediately obtain the ground state energy for the 2D-RVB state:
\[
F(\theta=0)=6J_{1}\xi^{2}+12J_{2}\eta^{2}-\intop g(\varepsilon)\varepsilon d\varepsilon\]
 The integral is done analytically \cite{Janke-Emde_Loesch}: \[
\frac{6\sqrt{2}}{\pi}\sqrt{J_{1}^{2}\xi^{2}+J_{2}^{2}\eta^{2}}\mathsf{E}\left(k\right)\]
 where $\mathsf{E}(k)$ is the complete elliptic integral of the second
kind of the modulus $k$ given by:\[
k^{2}=\frac{J_{1}^{2}\xi^{2}}{J_{1}^{2}\xi^{2}+J_{2}^{2}\eta^{2}}.\]
This result is not unexpected since it has a form characteristic for
one-dimensional systems \cite{Fulde-book}. Taking derivatives with
respect to the order parameters and setting them equal to zero results
in self-consistency conditions: \begin{eqnarray}
1 & = & \frac{J_{1}}{\sqrt{2}\pi\sqrt{J_{1}^{2}\xi^{2}+J_{2}^{2}\eta^{2}}}\left(\mathsf{K}\left(k\right)-\mathsf{D}\left(k\right)\right)\label{eq:Q1DSelfConsistency}\\
1 & = & \frac{J_{2}}{2\sqrt{2}\pi\sqrt{J_{1}^{2}\xi^{2}+J_{2}^{2}\eta^{2}}}\mathsf{K}\left(k\right)\nonumber \end{eqnarray}
which are remarkably similar to the self-consistency conditions \cite{Tch010}
in the one-dimensional Hubbard problem, the first being one for the
\emph{bond order} and the second being analogous to that for the \emph{gap}
or \emph{magnetization} with the $J_{2}$ parameter taking part of
the interaction parameter $U$ of the Hubbard model and $3\sqrt{2}J_{1}\xi$
being the effective one-dimensional bandwidth. In the 1D-RVB state
the first of the conditions eq. (\ref{eq:Q1DSelfConsistency}) yields
the amplitude of the order parameter $\xi$ reached at the zero temperature:\begin{equation}
\xi_{0}=\frac{1}{\sqrt{2}\pi},\label{eq:LimitOrderParameter}\end{equation}
which is in perfect agreement with the numerical result of \cite{Hayashi and Ogata }.
Inserting this in the second of the two conditions eq. (\ref{eq:Q1DSelfConsistency}),
neglecting the terms with containing $\lyxmathsym{\textgreek{h}}$
as compared to those with \textgreek{x} in the sums, and using the
logarithmic asymptotic form of the complete elliptic integral of the
first kind $\mathsf{K}$ we arrive to the self-consistent field-like
estimate of the for $\eta$ and for the gap in the 2D-RVB state: \begin{eqnarray*}
\eta & = & \frac{2\sqrt{2}}{\pi}\frac{J_{1}}{J_{2}}\exp\left(-\frac{2J_{1}}{J_{2}}\right)\\
3\sqrt{2}J_{2}\eta & = & \frac{12J_{1}}{\pi}\exp\left(-\frac{2J_{1}}{J_{2}}\right)\end{eqnarray*}
at zero temperatue. The latter result is in a fair agreement with
the numerical study \cite{Hayashi&Ogata-arxiv} where the pre-exponential
factor in the gap was estimated to be 3.50 as compared to 12/\textgreek{p}
\ensuremath{\approx} 3.82. Although the factor in the exponent was
estimated to be 1.61 in \cite{Hayashi&Ogata-arxiv} against our estimate
of two, the general form of the dependence of the characteristics
of the model on its parameters is reproduced. These results in agreement
with the numerical results \cite{Hayashi&Ogata-arxiv} and general
behavior of one-dimensional models with interaction and show that
at the zero temperature some nonvanishing value of the $\eta$ order
parameter and the energy gap appear at arbitrarily weak interaction
$J_{2}$ so that no critical point with respect to the anisotropy
$J_{2}/J_{1}$ should be expected at zero temperature. 

A further move consists in inserting the above expression for $\eta$
in the logarithm and retaining the terms proportional to $\Lambda k'^{2};\Lambda=\ln\frac{4}{k'};k\mathit{'}^{2}=1-k^{2}$
in the equation for $\xi$ so that $\mathsf{K}\left(k\right)-\mathsf{D}\left(k\right)\approx1-\Lambda k'^{2}/2$.
By doing so and retaining the terms up to second order in $\eta$
we obtain:\begin{equation}
\xi=\frac{1}{\sqrt{2}\pi}-\frac{\sqrt{2}\pi J_{2}}{J_{1}}\eta^{2},\label{eq:XivsEta}\end{equation}
which represents the estimate of the bond order variation in the 2D-RVB
state as compared to the 1D-RVB state. 

This finding is in a fair and remarkable agreement with the numerical
result of \cite{Hayashi&Ogata-arxiv} where it was shown that in the
region where the 2D-RVB state develops ($\eta\neq0$) the $\xi$ parameter
manifests a very weak depletion as compared to its 1D-RVB ($\eta=0$)
value. Despite the fact that it must be not particularly strong this
depletion can manifest itself in a geometry change, which in principle
could be observed. The subsequent reasoning following the lines proposed
in \cite{MisurkinOvchinnikov} shows how it can tentatively look like. 

Assume that the lattice contribution to the total energy per copper
site can be harmonically approximated as a function of the separation
$\lyxmathsym{\textgreek{r}}$ between the Cu atoms in the \emph{a}-direction
(the shortest such separation in the structure): $K(\text{\textgreek{r}}-\text{\textgreek{r}}_{0})^{2}/2$,
where $K$ is an effective elastic constant and $\lyxmathsym{\textgreek{r}}_{0}$
is an equilibrium separation to be observed if the spin contribution
eqs. (\ref{eq:FreeEnergy}),(\ref{eq:Q1DFreeEnergy}) to the energy
are turned off. The spin contribution to the lowest order comes from
the \textquotedblleft{}kinetic energy\textquotedblright{} term and
is given by $-(6\lyxmathsym{\textsurd}2)/\lyxmathsym{\textgreek{p}}J_{1}\xi$.
Assuming the dependence of the effective exchange integral on the
interatomic separation in the form$J_{1}=J_{10}+J_{1}^{\prime}(\lyxmathsym{\textgreek{r}}-\lyxmathsym{\textgreek{r}}_{0})$
we easily arrive to the analog of the famous bond order \emph{vs}.
bond length relation \cite{CoulsonGolebievski} for the RVB states: 

\begin{equation}
\lyxmathsym{\textgreek{r}}-\lyxmathsym{\textgreek{r}}_{0}=\frac{6\sqrt{2}}{\lyxmathsym{\textgreek{p}}}(J_{1}^{\prime}\xi)/K,\label{eq:RhovsXi}\end{equation}
where $\lyxmathsym{\textgreek{r}}$ is now the equilibrium interatomic
separation in the presence of the spin contribution to the energy.
Obviously, the separation is going to change according to the sign
of the derivative of the effective exchange integral with respect
to the interatomic separation increase ($J_{1}^{\prime}<0$). The
latter condition is, however, natural, in the frame of the standard
conception of the sources of the antiferromagnetic exchange. Indeed,
these appear as a result of perturbative treatment of the one-electron
hopping in a strongly interacting regime. Then one has: $J_{i}=(4t_{i}^{2})/U$
where $t_{i}$ is the intersite one-electron hopping parameter along
the respective hopping vector $\boldsymbol{\tau}_{i}$ and $U$ is
the on-site electron-electron repulsion parameter. Assuming a linear
dependence of the hopping parameter on the interatomic separation:
$t=t_{0}+t\mathit{'}(\lyxmathsym{\textgreek{r}}-\lyxmathsym{\textgreek{r}}_{0})$
we arrive to an estimate $J_{1}^{\prime}=(8t_{0}t\mathit{'})/U<0$
since one can easily see that the multipliers $t_{0}$ and $t\mathit{'}$
must have opposite signs. Combining eqs. (\ref{eq:XivsEta}) and (\ref{eq:RhovsXi})
we arrive to the estimate for the variation of the equilibrium interatomic
separation in the 2D-RVB state: 

\[
\lyxmathsym{\textgreek{d}\textgreek{r}}=\text{\textendash}6J_{2}J_{1}^{\prime}\eta^{2}/J_{1}K\]
which immediately shows that the lattice parameter $a$ in CuNCN must
manifest the same trend as the 2D-RVB gap ($J_{1}^{\prime}<0$), although
with somewhat damped amplitude due to square in a small quantity $\eta$.

\section{Conclusion}

In the present paper we succeeded in obtaining analytical estimates
for the parameters (order parameters, energy gap) of the 2D-RVB state
and possible structural concequences of the variation of these parameters
under a 1D to 2D-RVB transition conjectured recently \cite{10SmallNegroes}
to be responsible for the observed temperature behavior of the magnetic
susceptibility in CuNCN.

\section*{Acknowledgments}

This work has been performed with the support of Deutsche Forschungsgemeinschaft.
We acknowledge the Russian Foundation for Basic Research for the financial
support dispatched to ALT through the grant No. 10-03-00155.

\end{document}